\documentstyle[manuscript,aps]{revtex}
\begin{document}
\draft
\title{A possible test for quadratic gravity in $d\geq 4$
dimensions} 
\author{Janusz Garecki}
\address{Institute of Physics, University of Szczecin, Wielkopolska 15;
70-451 Szczecin, POLAND}
\date{\today}
\maketitle
\begin{abstract}
In the letter we consider the Einsteinian strengths and dynamical degrees
of freedom for quadratic gravity. We show that the purely metric quadratic
gravity theories are much more stronger in Einsteinian sense than
the competitive quadratic gravity theories which admit torsion.
\end{abstract}
\newpage
\section{Introduction}
The notion of ``strength of the field equations'' was introduced in past
by Einstein [1] in order to analyze systems od partial differential
equations for physical fields. Later this notion was examined and
effectively used in field theory by several authors [2--6]. In
particular Schutz [3] pointed out that that the Einsteinian strength of the
field equations is closely  connected with the {\it number of the dynamical
degrees of freedom} which these equations admit in {\it Cauchy problem}.

In this letter we analyze the Einsteinian strength and related numbers
for a typical purely metric 4th-order gravity [7--14] which follows from the
Lagrangian 
\begin{equation}
L_g = \chi R + c_0 R^2 + c_1\vert Ric\vert^2 + c_2\vert Riem\vert^2,
\end{equation}
where $\chi,~~c_0,~~c_1,~~c_2$ are some dimensionals constants,
and for a typical ``Poincar\'e gauge quadratic field theory of
gravity'' ({\bf PGT}) with torsion [15--20].

The gravitational Lagrangian $L_g$ for {\bf PGT} can only be quadratic
in curvature like (1) (but admitting torsion) or can contains terms
quadratic in curvature (like (1)) plus terms quadratic in irreducible
components of torsion.\footnote{Concerning the most general Lagrangian
$L_g$ for {\bf PGT} see e.g. [16].}

It is commonly known that one can always get a symmetric
energy--momentum tensor for matter $T^{ik} = T^{ki}$ starting from the {\it canonical
pair} 
\begin{equation}
_cT^{ik}, ~~_c S^{ikl} = (-) _c S^{kil}:~_c T^{ik} - _c T^{ki} =\nabla_l
{}_c S^{ikl}.
\end{equation}
$_c T^{ik}\not= _c T^{ki}$ means here a canonical energy--momentum
tensor for matter and $_c S^{ikl} = (-) _c S^{kil}$ its canonical
spintensor (see e.g. [22]). It can be easily done by use  of the {\it
Belinfante symmetrization procedure} [21,22]. The symmetric
energy--momentum tensor $T^{ik}$ gives at least as well description of the
energy--momentum and angular momentum of matter as the canonical pair $(_c
T^{ik}, ~~_c S^{ikl})$ gives; but it is simpler and  has more symmetry and
better conservative properties. 

To geometrize the symmetric energy--momentum tensor $T^{ik} = T^{ki}$
the metric $g_{ik}$ (and Levi--Civita connection) is sufficient, i.e.,
the (pseudo)- Riemannian geometry is sufficient. A geometrization of
such a kind leads us to general relativity ({\bf GR}) and to its quadratic
(higher--order) purely metric generalizations. The field equations are here obtained
by use {\it Hilbert} (or metric) {\it variational principle} and have the
following general form 
\begin{equation}
{\delta\sqrt{\vert g\vert} L_g\over\delta g^{ab}} = {\delta\sqrt{\vert
g\vert} L_{mat}\over\delta g^{ab}}~ (=
1/2 \sqrt{\vert g\vert}T_{ab}).
\end{equation}
The ten field equations (3) are, in general, of the 4th--order.

In order to geometrize the canonical pair $(_c T^{ik}, ~~_c S^{ikl})$
the {\it Palatini variational principle} and a something more general
metric geometry, namely {\it Riemann--Cartan} geometry with torsion are
needed. 

In the Palatini variational principle we take $g_{ik},~\Gamma^i_{~kl}$
as independent variables (or, equivalently, an orthonormal tetrad
$h^{(a)}_{~~~i}(x)$ and ``Lorentz connection''
$\Gamma_{(i)(k)}^{~~~~~~l}$ [14],[20]). This variational principle leads us to Einstein--Cartan--Sciama--Kibble ({\bf ECSK})
theory of gravity and to its generalizations ---- Poincar\'e gauge quadratic
field theories of gravity ({\bf PGT}). The forty field equations are here of
the 2nd--order and they have the following general form \footnote{If we
take metric $g_{ik}$ and connection $\Gamma^i_{~kl}$ as independent
variables.} 
\begin{eqnarray}
{\delta\sqrt{\vert g\vert} L_g\over\delta g^{ab}} & = &
{\delta\sqrt{\vert g\vert} L_{mat}\over\delta
g^{ab}} (= \sqrt{\vert g\vert} _c T_{(ab)}),\nonumber \\
{\delta\sqrt{vert g\vert} L_g\over\delta\Gamma^{ik}_{~~l}}& = & {\delta
\sqrt{\vert g\vert}L_{mat}\over\delta\Gamma^{ik}_{~~l}}~(= \sqrt{\vert
g\vert} _c S_{ik}^{~~l}) 
\end{eqnarray}
plus additional metricity constraints
\begin{equation}
\nabla_i g_{kl} = 0.
\end{equation}
The antisymmetric part $_c T_{[ab]}$ of the canonical energy--momentum
is determined by covariant divergence of the canonical spintensor $_c
S^{ikl} = (-) _c S^{ikl}$ and by vectorial part of torsion (see e.g.
[6],[20]). 
The field equations (4)-(5) are of the 2nd-order with respect to
$g_{ik}$ and $\Gamma^i_{~kl}$ or, equivalently, they are of the 3rd-
order with respect to the really independent variables: metric and
contorsion (see e.g. [6,16]). 

It is remarkable that the both initial theories in these two
geometrization schemes ---- {\bf GR} and {\bf ECSK} theory of gravity
---- {\it have the same Einsteinian strengths} (12 in four dimensions) and
admit {\it the same numbers dynamical degrees of freedom} (4 in four dimensions)
in Cauchy problem. But, as we will see, the pure metric scheme to
geometrize symmetric energy--momentum tensor of matter leads us to the
quadratic gravity theories (in general of 4th--order) which have much
more smaller strenghts (48 in four dimensions) and numbers dynamical
degrees of freedom (16 in four dimensions) than the competitive {\bf
PGT} (120 and 40 in four dimensions respectively).

Thus, following Einstein [1] the purely metric geometrization scheme gives
us {\it much more stronger}, i.e., {\it better} from the physical point
of view  field equations\footnote{If strenght is
{\it smaller} then the corresponding field equations are {\it stronger},i.e.,
then the field equations {\it more precisely determine physical field}.}
than the competitive {\bf PGT}. This fact can be used as a {\it possible
test} for quadratic gravity. Namely, following Einstein one should
choose the purely metric quadratic gravity theories as {\it the better ones}
from the all set of the quadratic gravity theories.

In general, one can easily see that the Palatini variational principle leads
us to the field equations (of the 2nd-order but much more greater in number) which {\it are
not equivalent} for the same Lagrangian $L_g$ to the ten purely metric
field equations obtained by use of the Hilbert variational principle
(exception is the general relativity Lagrangian $L_g = \chi R$). \footnote{For
extended discussion of this problem see [14].}

The Einsteinian strengths $S_E(d)$ and numbers of the dynamical degrees
of freedom $N_{DF}(d)$ for the field equations obtained by Palatini
variational principle are much more greater then the corresponding
quantities for the purely metric gravity theories obtained by use
Hilbert variational principle. This means that the Palatini variational
principle gives {\it much more weaker}, i.e., {\it worse} from the
physical point of view  gravitational field equations
than the Hilbert variational principle. Only the so--called
``constrained Palatini variational principle'' [14] with Lagrange
multipliers gives gravitational field equations which are fully equivalent
to that obtained by use of the Hilbert variational principle. 
\section{Strengths and dynamical degrees of freedom for quadratic
gravity theories in $d\geq 4$ dimensions}
Using the definitions  and formulas given in [1--6] one can very easy find the
following number $Z_n(d)$ of the free coefficients of order $n$ in Taylor's
expansion of an analytic solution for the pure metric quadratic theory of gravity with the
Lagrangian (1) in the general case
\begin{eqnarray}
Z_n(d)& =& {d(d+1)\over 2}\left [\matrix{d\cr
n\cr}\right] - d\left [\matrix{d\cr
n+1\cr}\right] - {d(d+1)\over 2}\left[\matrix{d\cr
n-4\cr}\right] \nonumber \\
&+& d\left[\matrix{d\cr
n-5\cr}\right]\asymp \left[\matrix{d\cr
n\cr}\right]{2d(d-1)(d-2)\over n}\asymp 2d(d-2)\left[\matrix{d-1\cr
n\cr}\right],
\end{eqnarray}
where
\begin{equation}
\left[\matrix{d\cr
n\cr}\right] := {(n+d-1)!\over n!(d-1)!}.
\end{equation}
The symbol $\asymp$ means equality in the highest powers of $n$.
$n\longrightarrow\infty$.  

In the formula (6) the first term on the right gives the total number of
the nth--order coefficients and the other terms, before the sign
$\asymp$, give numbers of independent conditions imposed on these
nth--order coefficients: $d\left[\matrix{d\cr
n+1\cr}\right]$ conditions follows from {\it gauge freedom} and
$\Biggl\{\left(\matrix{d+1\cr
2\cr}\right)\left[\matrix{d\cr
n-4\cr}\right] - d\left[\matrix{d\cr
n-5\cr}\right]\Biggr\}$ conditions follow from {\it vacuum field equations}
and from {\it differential identities} which are satisfied by them (see
e.g. [1--6]).

One can easily read from these expressions that the {\it Einsteinian
strength} $S_E(d)$\footnote{$S_E(d)$ is defined as the coefficient of $1/n$
in the ratio $Z_n(d)/{d\brack
n}$.}  for such a theory is equal 
\begin{equation}
S_E(d) = 2d(d-1)(d-2)
\end{equation}
and that the {\it number dynamical degrees of freedom} $N_{DF}(d)$
\footnote{$N_{DF}(d)$ is the limit for large $n$ of
$Z_n(d)/{d-1\brack
n}$. $N_{DF}(d)$ is the number of free functions of $(d-1)$
variables in the theory (see 
e.g. [3]).} equals
\begin{equation}
N_{DF} (d) = 2d(d-2).
\end{equation}

The formula (6) can be expanded in the other form proposed by Schutz [3]
\begin{eqnarray}
Z_n(d) &=& 2d(d-2)\left[\matrix{d-1\cr
n\cr}\right] + \bigl(9d -3d(d+1)\bigr)\left[\matrix{d-2\cr
n\cr}\right]\nonumber \\
&+& \bigl(2d(d+1) - 11d\bigr)\left[\matrix{d-3\cr
n\cr}\right] + \bigl(4d - {d(d+1)\over 2}\bigr)\left[\matrix{d-4\cr
n\cr}\right]\nonumber \\
&-& 2d\left[\matrix{d-5\cr
n\cr}\right] -d\bigl(\sum_{k = d-6}^{d-(d-1)}\left[\matrix{k\cr
n\cr}\right]\bigr).
\end{eqnarray}

The physical meaning of the coefficients in the above
expansion, except of the first coefficient $2d(d-2) =: N_{DF}(d)$, is
unclear.\footnote{There were given trials to understand the physical
meaning of the rest coefficients of the $Z_n(d)$ (see e.g. [4,5]).} 

In four dimension ($d = 4$) we have from (6) or from (10) 
\begin{eqnarray}
Z_n(4)& =& 16\left[\matrix{3\cr
n\cr}\right] - 24\left[\matrix{2\cr
n\cr}\right] -4\left[\matrix{1\cr
n\cr}\right]\nonumber \\
&\asymp& \left[\matrix{4\cr
n\cr}\right]{48\over n} \asymp 16\left[\matrix{3\cr
n\cr}\right],
\end{eqnarray}
i.e., we have here the strength equal 48 and  16 degrees of
freedom.\footnote{In special cases these numbers can be smaller
[11,12,22].} 

We must emphasize that there exist an interesting example of the
quadratic and pure metric theory of gravity called the {\it
Einstein--Gauss--Bonnet} theory [22] ({\bf EGB}) which has specific
gravitational Lagrangian $L_g$ of the form 
\begin{eqnarray}
L_g &=& L_E + L_{GB} = L_E \nonumber \\
&+& \alpha\bigl(R_{iklm} R^{iklm} - 4 R_{ik} R^{ik} + R^2\bigr),
\end{eqnarray}
where $\alpha$ is a new coupling constant.

The Lagrangian $L_{GB}$ is called {\it Gauss--Bonnet} or {\it Lovelock}
Lagrangian.  

The field equations of this theory are of the 2nd--order for $d\geq
4$ (iff $d = 4$, then these field equations are simply Einstein
equations) and they have the same strength $S_E(d)$ and number dynamical
degrees of freedom $N_{DF}(d)$ as Einstein equations have, i.e., they
have 
\begin{equation}
S_E(d) = d(d-1)(d-3),~~N_{DF}(d) = d(d-3).
\end{equation}
Iff $d = 4$ we have for this theory 
\begin{equation}
S_E(4) = 12, ~~N_{DF}(4) = 4.
\end{equation}
We see that the Lagrangian (12) leads us to {\it the strongest} field
equations. Moreover the quadratic theory of gravity with Lagrangian (12) {\it
admits no ghosts or tachyons} in its linear approximation. 

Following Einstein [1] the {\bf EGB} quadratic theory of
gravity is the {\it best one theory} from the all set of the purely metric
gravity theories which have quadratic gravitational Lagrangian of the
general form (1). It is because the {\bf EGB} theory of gravity has  the
strongest field equations.

On the other hand for a standard {\bf PGT} \footnote{With or without
terms quadratic in torsion in its Lagrangian} we have [6,26]
\begin{eqnarray}
Z_n(d) &=& {d(d+1)\over 2}\left[\matrix{d\cr
n\cr}\right] + {d(d-1)\over 2}d\left[\matrix{d\cr
n-1\cr}\right] - d\left[\matrix{d\cr
n+1\cr}\right]\nonumber \\
&-& \Bigl\{d^2\left[\matrix{d\cr
n-2\cr}\right] + {d(d-1)\over 2}d\left[\matrix{d\cr
n-3\cr}\right] - {d(d-1)\over 2}\left[\matrix{d\cr
n-4\cr}\right]\nonumber \\
&-& d\left[\matrix{d\cr
n-3\cr}\right]\Bigr\}\asymp \left[\matrix{d\cr
n\cr}\right]{d(d+1)(d-1)(d-2)\over n}\asymp
d(d+1)(d-2)\left[\matrix{d-1\cr
n\cr}\right].
\end{eqnarray}
This gives for $d = 4$ 
\begin{equation}
Z_n(4)\asymp \left[\matrix{4\cr
n\cr}\right]{120\over n}\asymp 40\left[\matrix{3\cr
n\cr}\right],
\end{equation}
i.e., we have here $S_E(d) = 120, ~~N_{DF}(d) = 40$.

In the formula (15), likely as it was in the formula (6), the first two
terms on the right give the total number of the nth--order coefficients
and the other terms before the sign $\asymp$ give numbers of independent
conditions imposed on these nth--order coefficients: $ d\left[\matrix{d\cr
n+1\cr}\right]$ follows from gauge freedom and
$\Biggl\{d^2\left[\matrix{d\cr
n-2\cr}\right] + d\left(\matrix{d\cr
2\cr}\right)\left[\matrix{d\cr
n-3\cr}\right] - \left(\matrix{d\cr
2\cr}\right)\left[\matrix{d\cr
n-4\cr}\right] -d\left[\matrix{d\cr
n-3\cr}\right]\Biggr\}$ conditions follow from the vacuum field equations
and from differential identities which are satisfied by them (see e.g.
[6]). 

Comparing (11) and (16) we see that a typical 3rd--order {\bf PGT} has in
four dimensions almost {\it three times greater strength and number dynamical
degrees of freedom} than a typical 4th--order purely metric quadratic
gravity theory.

Note also that the formal limes
\begin{equation}
\displaystyle\lim_{d\to\infty}{Z_n^{PGT}\over Z_n^{FOTH}} =\infty,
\end{equation}
i.e., it is infinite.
This means that if $d$ is growing than the field equations of the purely
metric quadratic gravity becomes more and more stronger in comparison
with the field equations of a {\bf PGT}.
\section{Conclusion}
In the letter we have considered Einsteinian strengths $S_E(d)$ and
number dynamical degrees of freedom $N_{DF}(d)$ for a typical 4th--order purely metric theory of gravity with
general gravitational Lagrangian of the form (1). Such theory of gravity gives us
a typical example of a purely  metric, quadratic theory of gravity. We
have compared these numbers $S_E(d)$ and $N_{DF}(d)$ with the analogous numbers
for a typical {\bf PGT} with torsion. As we have seen, the numbers
$S_E(d)$ and $N_{DF}(d)$ for a typical {\bf PGT} are much more greater than for
a pure quadratic metric gravity. This means that the purely metric
quadratic gravity theories obtained by use Hilbert variational principle
have much more stronger field equations than the competitive quadratic
theories of gravity with torsion obtained by use Palatini variational
principle. Following Einstein [1], if one have no other criterion, one
choose as the better this theory of gravity, which has stronger field
equations. So, following Einstein, one should treat the purely metric
quadratic theories of gravity  as {\it the better} quadratic theories of
gravity than the competitive quadratic theories of gravity with torsion.

Among these purely metric quadratic theories of gravity in $d\geq 4$ the
2nd--order {\bf EGB} theory {\it has the strongest field equations},
i.e., this is the {\it best one theory} from the all set of the purely metric
quadratic theories of gravity.
\centerline{\bf Acknowledgements}

The author would like to think Prof. Rainer Schimming for suggesting
this research project and for his very useful hints and comments.

The author would like also to thank the EMA  University in Greifswald and
the Max-Planck-Institut f\"ur Gravitationsphysik in Golm where most of
this work has been done for warm hospitality. Especially, he would like to
thank DAAD, Bonn,  for financial support.

\end{document}